\documentclass[conference]{IEEEtran}
\IEEEoverridecommandlockouts
\usepackage{cite}
\usepackage{amsmath,amssymb,amsfonts}
\usepackage{algorithmic}
\usepackage{graphicx}
\usepackage{textcomp}
\usepackage{xcolor}

\usepackage{epsfig}

\def\BibTeX{{\rm B\kern-.05em{\sc i\kern-.025em b}\kern-.08em
    T\kern-.1667em\lower.7ex\hbox{E}\kern-.125emX}}
\begin{document}

\title{Learning Sentinel-2 Spectral Dynamics for Long-run Predictions using Residual Neural Networks\\
\thanks{This work has been supported by the Programme National de Télédétection Spatiale (PNTS), grant n$^\circ$PNTS-2019-4.}}

\author{\IEEEauthorblockN{Joaquim Estopinan}
\IEEEauthorblockA{\textit{EPITA Research and Development} \\ \textit{Laboratory (LRDE)} \\
Le Kremlin-Bicêtre, France \\
joaquim.estopinan@inria.fr}
\and
\IEEEauthorblockN{Guillaume Tochon}
\IEEEauthorblockA{\textit{EPITA Research and Development} \\ \textit{Laboratory (LRDE)} \\
Le Kremlin-Bicêtre, France \\
guillaume.tochon@lrde.epita.fr}
\and
\IEEEauthorblockN{Lucas Drumetz}
\IEEEauthorblockA{\textit{Lab-STICC, UMR CNRS 6285} \\
\textit{IMT Atlantique}\\
Brest, France \\
lucas.drumetz@imt-atlantique.fr}
}

\maketitle

\begin{abstract}
Making the most of multispectral image time-series is a promising but still relatively under-explored research direction because of the complexity of jointly analyzing spatial, spectral and temporal information. Capturing and characterizing temporal dynamics is one of the important and challenging issues. Our new method paves the way to capture real data dynamics and should eventually benefit applications like unmixing or classification.
Dealing with time-series dynamics classically requires the knowledge of a dynamical model and an observation model.
The former may be incorrect or computationally hard to handle, thus motivating data-driven strategies aiming at learning dynamics directly from data.
In this paper, we adapt neural network architectures to learn periodic dynamics of both simulated and real multispectral time-series. We emphasize the necessity of choosing the right state variable to capture periodic dynamics and show that our models can reproduce the average seasonal dynamics of vegetation using only one year of training data.
\end{abstract}


\section{Introduction} 
\label{sec:intro}

The Sentinel-2 mission is part of the European Earth observation project Copernicus, a program jointly led by the European Space Agency (ESA) and the European Commission. It intends to provide authorities and interested actors with open multispectral data reflecting Earth surface changes, with proper natural resources management as end goal.
The mission consists in two polar-orbiting satellites synchronized on the same sun orbit, and diametrically opposite to one another. As each Sentinel-2 satellite has a temporal revisit of ten days, the common temporal revisit of a given location on earth under the same viewing conditions goes down to five days~\cite{S2}.\\
The Sentinel-2 mission (or the soon to be launched pending hyperspectral EnMAP mission~\cite{guanter2015enmap}) gives access to spectral time-series with a high temporal revisit. Such multidimensional data require a particular care to extract their rich information, but proved to greatly increase carried out task results when processed successfully \cite{he2018multi,arch_sites,yokoya2017multisensor}.
The automatic learning of spectral dynamics would provide a new spectral insight on the observed areas. It could benefit a large panel of applications such as spectral unmixing, anomaly detection, data assimilation~\cite{evensen2009data} and predictions on vulnerable species conservation status~\cite{EFTs}. It was proven in \cite{learning_em_dyn} that knowing spectral dynamics improves an endmember estimation task on synthetic data.\\
In the observed scenes, pure materials of interest are varying through time in their spectral signature or/and their spatial extent because of seasonal changes for instance. Reference ~\cite{S_Henrot_SU_overview} showed that a state-space model is convenient to model these evolutions. For a given material described by a $n$-dimensional state variable $\mathbf{X}_{t}$, the \textit{state equation} of the model is given by an ordinary differential equation (ODE) when only temporal variations of the spectra are considered:
\begin{equation}\label{eq:ODE}
     \frac {d\mathbf{X}_{t}} {dt} = \mathbf{\mathcal{F}}(\mathbf{X}_{t})+\mathbf{\eta}_{t}
\end{equation}
where $\mathbf{\mathcal{F}}$ is the dynamical operator and $\mathbf{\eta}_t$ an error accounting for both noise and model approximation. Our objective is thereby to directly learn from data these dynamics with specific neural network (NN) architectures~\cite{learning_em_dyn}.\\
The contribution of this article is threefold: \textit{(i)} a particular state variable describing spectral signatures is shown efficient to learn periodic dynamics with residual neural networks (ResNets), \textit{(ii)} long-run predictions on synthetic data show that the developed architecture is clearly more adapted for such task than a long short-term memory (LSTM)~\cite{Hochreiter1997LSTM} architecture, \textit{(iii)} first results on real data demonstrate the potential of our method for applications like classification or unmixing.\\
In the following, Section~\ref{sec:dynamics} focuses on methods and NNs to learn spectral dynamics. Then, Sections~\ref{sec:exp} and~\ref{sec:results} introduce conducted experiments and results on synthetic and real Sentinel-2 data, respectively. Finally, Section~\ref{sec:discussion} gathers our conclusions on this work and its implications.

\begin{figure*}[tb]
\begin{minipage}[b]{1.0\linewidth}
  \centering
  \centerline{\epsfig{figure=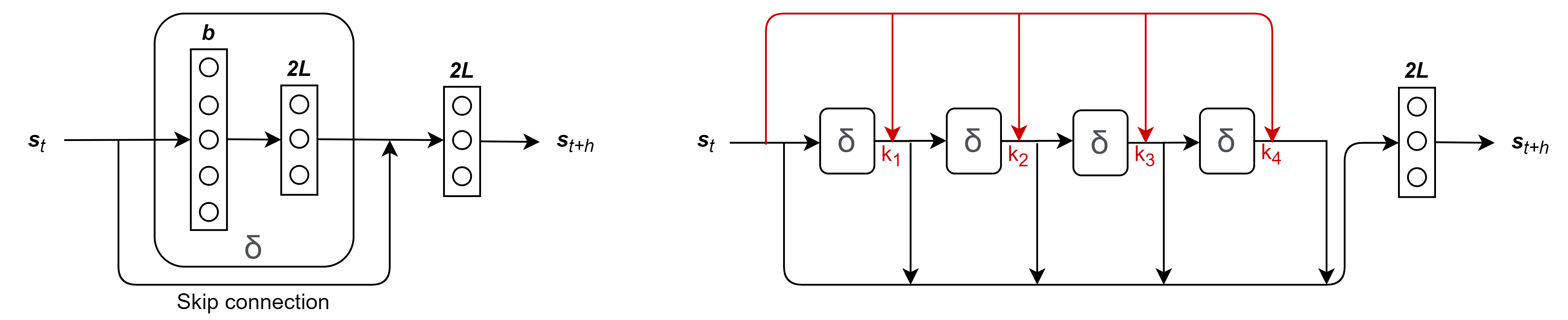,width=17cm}}
  \centerline{(a) Euler \hspace{7cm} (b) Runge-Kutta 4}\medskip
\end{minipage}
\caption{ResNets architectures implementing (a) Euler and (b) Runge-Kutta integration schemes.}
\label{fig:resnets}
\end{figure*}


\section{Dynamics learning}
\label{sec:dynamics}

\subsection{ResNet architecture}
\label{ssec:Resnets}

Recurrent neural networks (RNNs) architectures, such as LSTM networks~\cite{Hochreiter1997LSTM}, are natural candidates when it comes to learning and processing time-series data. Among RNNs architectures, ResNets with shared weights across layers have recently been proven to be particularly adapted to estimate dynamical operators~\cite{bilin_res}. Using them allows to reformulate the problem of learning an input/output relationship to the learning of a deviation from the identity~\cite{chen2019neural, bilin_res}, making them suited to process data generated by differential equations. Mathematically, we assume that spectral signatures can be obtained by the following equation inherited from the integration of the ODE \eqref{eq:ODE}:
\begin{equation}\label{eq:s_t+1}
     \mathbf{s}_{t+h} = \mathbf{s}_{t} + h\mathbf{\delta(s}_{t}),
\end{equation}
$\mathbf{s}_{t}$ being the pixel $L$-dimensional reflectance at time-step $(t)$, $L$ is the number of exploited spectral bands, $\delta$ is the infinitesimal operator of the integrated ODE and $h$ is the integration step (herafter set to 1 without loss of generality). In the ResNet typical architecture, a hard-cabled \textit{skip connection} constrains the learning of the difference between $\textbf{s}_{t+h}$ and $\textbf{s}_t$, see Fig.~\ref{fig:resnets} (a). The problem is thus shifted to the learning of the infinitesimal operator $\delta$. To achieve this, two numerical integration procedures are classically considered: the first-order Euler method and the more refined fourth-order Runge-Kutta 4 (RK4) method. These numerical integration schemes are hard-coded in the architectures. The ResNets are then learning the ODE through the weights attributed to the infinitesimal operator $\delta$.
         
\subsection{Euler and Runge-Kutta 4 integration schemes}
\label{ssec:int_schemes}
The classical first-order Euler method consists in approximating $\delta$ by the temporal derivative of the targeted function:
\begin{equation}\label{eq:euler}
     \mathbf{s}_{t+1} = \mathbf{s}_{t} + \mathbf{s}'_{t}
\end{equation}
An approximation of $\mathbf{s}'_{t}$ is therefore learned through the shared weights of $\delta$, see Fig.~\ref{fig:resnets} (a) (the rationale for the dimension $2L$ is further explained in Section~\ref{ssec:AS}). The two layers encoding $\delta$ are distributed through the whole time-series. 
The local error, \textit{i.e.} the error per step, is on the order $O(h^2)$~\cite{hairer2006numerical}.\\
The architecture of the RK4 integration scheme is displayed by Fig.~\ref{fig:resnets} (b). Here, the approximation of $\delta$ is given by a weighted average of four slopes taken at the beginning, the middle and the end of the integration step (their respective weighting coefficients being $k_1, k_2, k_3$ and $k_4$), and the four $\delta$ blocks are put in series.
$\delta$ is still learned by the weights of two consecutive fully connected layers. $b$, the number of connections between them, is of particular interest as it regulates the complexity given to the encoding of $\delta$, and thus its expressiveness degree. Finally, the ResNets have to be provided with enough input information to accurately learn the dynamical operator's nature.

\subsection{State augmentation with derivatives}
\label{ssec:AS}

By modifying the initial state of the input data, \textit{i.e.} more concretely adding or removing features to the $n$-dimensional vectors in entry, the RNNs can lead to very different predictions.
Initially, $L$-dimensional reflectance vectors $\mathbf{s}_t$  were used as state variable. However, with this configuration, the tested RNNs were not able to capture the periodicity of simulated spectral dynamics. This was because the complete state of the $2^{nd}$ order system was never observed as the derivatives were missing~\cite{latent_dyn}. Thus, the state variable was augmented with reflectance left derivatives $\mathbf{s}'_t$:
\begin{equation}\label{eq:state_augm}
     \mathbf{X}_{t} \leftarrow \begin{bmatrix}
\mathbf{s}_{t} \\
\mathbf{s}^\prime_{t}
\end{bmatrix}
 \in \mathbb{R}^{2L},
\end{equation}
It explains the layers $2L$ output dimensions in Fig.~\ref{fig:resnets}. Choosing the right state variable allowed to learn dynamics periodicity on simulated data during the training, see Section~\ref{sec:exp}. If necessary, one can go one step further by adding higher-order derivatives to the state variable.

\subsection{Training and testing strategies}
\label{ssec:training_test_set}

\begin{figure}[tb]
\begin{minipage}[b]{1.0\linewidth}
  \centering
  \centerline{\epsfig{ figure=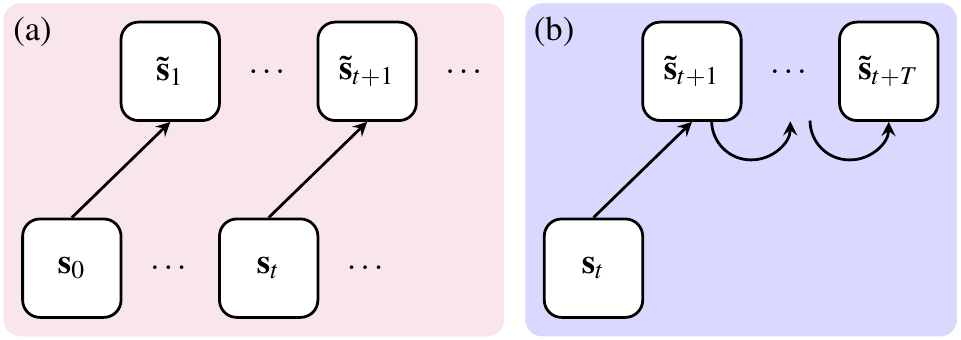, width=0.95\columnwidth}}
\end{minipage}
\caption{(a) Training on short-term predictions and (b) test on long-term predictions.}
\label{fig:short_long_term}
\end{figure}

RNNs are trained on successive one time-step prediction tasks, as depicted by Fig.~\ref{fig:short_long_term} (a). To predict $\tilde{\mathbf{s}}_{t}$, the RNNs are fed with the ground truth $\mathbf{s}_{t-1}$ at the previous date. Repeated along the training set, this process eventually enables to reach an interesting estimation of $\delta$. The weights of the two fully connected layers encoding the infinitesimal operator are updated for each short-term prediction done in the training set. Such one time-step predictions on reflectance values, introduced in~\cite{learning_em_dyn}, naturally flow from the the ResNets architecture implementing an integration scheme according to \eqref{eq:s_t+1}.\\
Now, the objective is to predict the reflectance of a pure pixel at time $(t+T)$ knowing its trajectory until date $(t)$ only. To do so, successive one time-step predictions are done from date $(t)$ up to date $(t+T)$. While the first prediction is indeed done from known data at time-step $(t)$, the RNN input is then fed with the prediction at the previous time-step until the date $(t+T)$ is reached, as shown by Fig.~\ref{fig:short_long_term} (b). It is a significantly harder task than the short-term predictions done during training since there is a higher chance that the overall error accumulates through the successive predictions, resulting in a trajectory drifting away from its expected behavior. Nevertheless, long-term predictions are useful to forecast dynamics ahead of time, and can be used to check that the learned dynamical systems has desirable topological properties (e.g. stability, convergence or even chaotic properties depending on the dynamics of the data~\cite{latent_dyn}).


\section{Validation on synthetic data}
\label{sec:exp}

\begin{figure}[tb]
\begin{minipage}[b]{1.0\linewidth}
  \centering
  \centerline{\epsfig{ figure=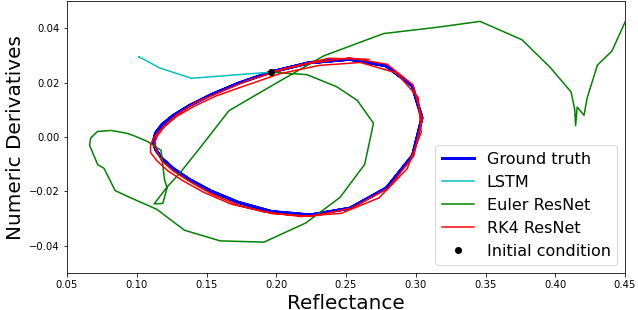, width=\columnwidth}}
\end{minipage}
\caption{Phase diagrams of LSTM, Euler ResNet and RK4 ResNet compared to the ground-truth trajectory. Represented long-run predictions are of 50 time-steps, $100^{th}$ spectral band.}
\label{fig:phase_diag}
\end{figure}

Simulated data were used to develop and test the different recurrent architectures~\cite{learning_em_dyn}. Time-series were constructed with hyperspectral ($L=144$) albedos derived from the DFC 2013 Houston dataset~\cite{albedos_article}, a daily-periodic cosine varying illumination angle and simplified Hapke model from~\cite{hapke_model}. We obtained a toy dataset following nonlinear dynamics. To learn these synthetic dynamics, the training was typically done on a twenty-day windows with a thirty-minute sample-step, 2500 epochs and \textit{Adam} optimizer.
Fig. \ref{fig:phase_diag} gathers phase diagrams of long-run prediction from LSTM, Euler ResNet and RK4 ResNet strategies compared to the ground truth ($100^{th}$ spectral band of the grass spectral signature).
RK4 ResNet clearly appears as the only network able to provide correct \emph{long-run} predictions (all architectures perform fine for 1 step-ahead predictions). LSTM and Euler ResNet do not manage to learn the slightly nonlinear daily-periodic variations, with predictions respectively collapsing and diverging after some oscillations. 
Table~\ref{tab1} presents the root-mean-square errors (RMSE) and spectral angle errors (SAE) for $(t+T)$ long-run predictions. RK4 ResNet predictions result in very low errors for first 120 time-steps (2.5 days). Then errors eventually begin to grow, which is reasonable given the iterative construction of long-run predictions (see Fig.~\ref{fig:short_long_term} (b)). 
Finally, it is worth mentioning that the synthetic dynamics are here the same for all spectral band, though with different albedo inputs for each band. However, this strong modeling hypothesis is likely not to be adapted to real data.

\begin{table}[t]
\caption{RMSE and SAE for LSTM and ResNets Long-run Predictions}
\begin{center}
    \begin{tabular}{|c|c|c|c||c|c|c|}
    \hline
    $\mathbf{t}$
    & \multicolumn{3}{c||}
    {\textbf{RMSE}$\mathbf{\times10^{-3}}$}
    & \multicolumn{3}{c|}
    {\textbf{SAE}$\mathbf{\times10^{-3}}$}
    \\
    \cline{2-7}
    $\mathbf{+}$   & \textbf{LSTM} & \textbf{Euler} & \textbf{RK 4} & \textbf{LSTM} & \textbf{Euler} & \textbf{RK 4}\\
    \hline
    $12$
    & $27.7$
    & $31.4$
    & $0.45$
    & $4.30$
    & $18.0$
    & $0.01$
    \\
    $24$
    & $48.5$
    & $32.9$
    & $3.01$
    & $4.30$
    & $13.0$
    & $0.01$
    \\
    $36$
    & $27.5$
    & $92.5$
    & $2.33$
    & $4.30$
    & $10.9$
    & $0.03$
    \\
    $120$
    & $49.8$
    & $2367$
    & $33.6$
    & $4.30$
    & $23.6$
    & $0.67$
    \\
    \hline
    \end{tabular}
    \label{tab1}
    \end{center}
\end{table}

\section{Experiments and results on real data}
\label{sec:results}

In this Section, we apply the developed spectral dynamic learning methodology on real Sentinel-2 time-series data.

\subsection{Data description}
\label{ssec:data_description}

\begin{figure*}[htb]

\begin{minipage}[b]{0.33\linewidth}
  \centering
  \centerline{\epsfig{figure=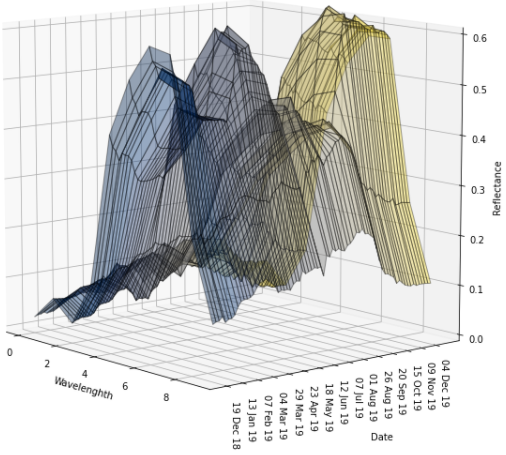,width=\columnwidth}}
  \centerline{(a) 2019 Training data}\medskip
\end{minipage}
\begin{minipage}[b]{0.33\linewidth}
  \centering
  \centerline{\epsfig{figure=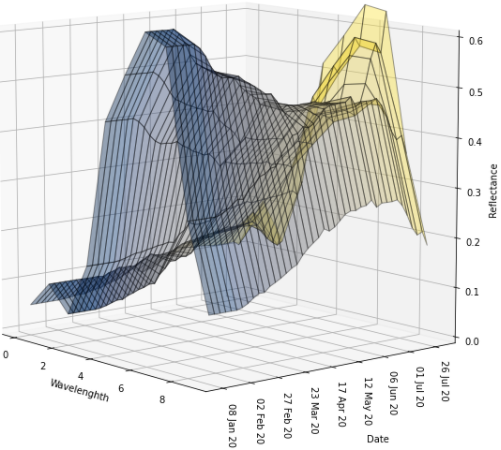,width=\columnwidth}}
  \centerline{(b) 2020 Ground truth}\medskip
\end{minipage}
\begin{minipage}[b]{0.33\linewidth}
  \centering
  \centerline{\epsfig{figure=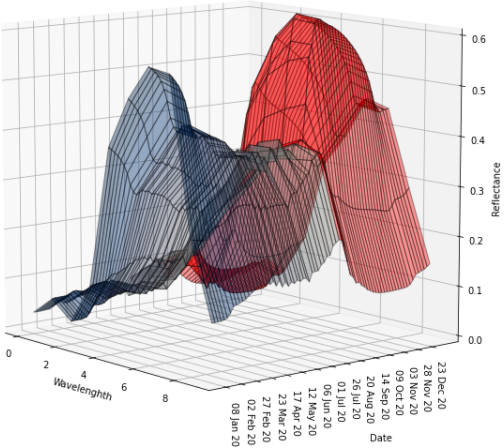,width=\columnwidth}}
  \centerline{(c) 2020 Long-run prediction}\medskip
\end{minipage}
\caption{Temporal evolution of a forest spectrum (a) in 2019 (whole year), (b) in 2020 (half year) and (c) 2020 long-run prediction (whole year) for the same pixel.}
\label{fig:3D_spectra}
\end{figure*}

\begin{figure}[tb]
\begin{minipage}[b]{1.0\linewidth}
  \centering
  \centerline{\epsfig{ figure=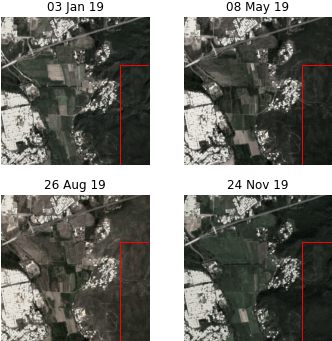, width=8cm}}   
\end{minipage}
\caption{Four RGB images extracted from the time-series.}
\label{fig:small_images}
\end{figure}

Images produced by Sentinel-2 satellite imaging sensors are composed of 4 bands with 10~m spatial resolution (covering visible and near infrared domains), 6 bands with 20~m spatial resolution (in near-infrared and short-wave infrared domains) and 3 bands with 60~m spatial resolution. Here, we only retained 10~m and 20~m bands ($L=10$), and upsampled those latter bands with bi-cubic interpolation to reach a common 10~m spatial resolution. 
The chosen area, located in South Puerto Rico, is the same as the one presented in~\cite{Guanica_article}, for which auxiliary data (such as rain records) have been collected at the close \textit{Ensenada} station.
The overall time-series runs from December 19$^{th}$ 2018 to July 31$^{st}$ 2020, resulting in a total of 118 individual images. Cloudy (thus unexploitable) images were replaced by Cressman interpolations \cite{cressman1959operational} conducted thanks to (timewise) nearby available data, with a Gaussian radius of interpolation set to $R = 10$ days. Finally, the complete time-series was globally normalized, \textit{i.e.} divided by the maximum reflectance value among all pixels, dates and spectral bands (the difference between 10~m and 20~m cells sensibility was also taken into account). RGB images extracted from the series are displayed in Fig.~\ref{fig:small_images}. 
We trained our model with forest pixels localized within the red rectangle. Forest indeed has a quite high intra-class spectral variability but the proposed approach is expected to learn an average spectral behavior. In the following, we only use the RK4 network since it proved to be the only one suited to long term predictions.

\subsection{Obtained results}
\label{ssec:obtained_results}
The training was done on 74 images ranging from December $19^{th}$ 2018 to December $19^{th}$ 2019. Fig.~\ref{fig:3D_spectra} (a) displays the spectral evolution of a given forest pixel along those dates. Data included in the training set for a targeted pixel comprised its four direct neighbors, leading to a batch size equal to 5. 60000 epochs were needed with $Adam$ optimizer and $MSE$ loss function to reach convergence and realistic results. 
The value given to $b$ turned out to be a major success criterion, the architecture being fixed otherwise. We found out that around 150 connections were optimal. 
Fig.~\ref{fig:3D_spectra} (b) represents the evolution of the same pixel as in Fig.~\ref{fig:3D_spectra} (a) between January $8^{th}$ 2020 and July $31^{st}$ 2020, serving as ground truth.
Finally, Fig.~\ref{fig:3D_spectra} (c) represents the long-run prediction over 73 dates (from January $8^{th}$ 2020 to January $12^{th}$ 2021) for this pixel. The red portion accounts for predictions after the $31^{st}$ of July 2020, to easily compare prior predictions with the ground truth. Predictions match the spectral decreasing tendency observable on the ground truth and peak prediction occurs slightly later than in reality (just after the $31^{st}$ of July). RK4 ResNet nonetheless achieves to learn specific and periodic patterns. The quality of the predictions decreases with time, but remains realistic. RMSE on January $18^{th}$ is of $5.0\times10^{-2}$ and of $7.3\times10^{-2}$ on April $17^{th}$. Because of the peak late prediction, it finally reaches $1.3\times10^{-1}$ on July $26^{th}$.
\section{Conclusion}
\label{sec:discussion}
In this work, we aimed at learning spectral dynamics patterns and periodicity directly from data, without prior knowledge. We showed that, using the right state variable, the developed ResNet implementing RK4 integration scheme succeeded in that task on simulated and real time-series multispectral data. Long-run predictions on synthetic data were proven significantly better than with LSTM or Euler ResNet, and realistic enough on a real multispectral time-series to capture seasonal variations in vegetation. Future research avenues include optimizing hyperparameters and adding an energy conservation constraint to further improve the results. A prior classification step on time-series to automatically integrate in training set spectrally consistent zones is also a promising direction, as is the stochastic modeling of the spectral variability of vegetation (prediction of a probability density instead of point estimates).
This work could eventually benefit applications for space time interpolation of multispectral data, scene unmixing and forecasting problems among many others. 

\clearpage
\bibliographystyle{IEEEbib}
\bibliography{conference_101719, strings}

\end{document}